\documentstyle{article}

\newcommand{\nc}{\newcommand}
\nc{\be}{\begin{equation}}
\nc{\ee}{\end{equation}}
\nc{\dfrac}{\displaystyle \frac }
\begin{document}

\author{ M. Arik$^{1}$, S. G%
\"{u}n$^{2}$, A. Yildiz$^{3}$ \\
\footnotesize $^{1}$ Department of Physics, Bo\={g}azi\c{c}i University, Bebek, Istanbul, Turkey
\\\
\footnotesize $^{2}$ Department of Physics, Istanbul University, Vezneciler, Istanbul,
Turkey \\
\footnotesize $^{3}$ Feza G\"ursey Institute, P.O. Box 6, 81220, \c Cengelk\"oy, Istanbul,
Turkey}
\title{\bf Invariance quantum group of the
fermionic oscillator }
\maketitle

\hspace {500pt}
\begin{abstract}

\baselineskip=20pt The fermionic oscillator defined by the algebraic relations $cc^{\ast
}+c^{\ast }c=1$ and $c^{2}=0$ admits the homogeneous group $O(2)$ as its
invariance group. We show that, the structure of the inhomogeneous invariance
group of this oscillator is a quantum group.
\end{abstract}



\hspace {150 cm}

\noindent \baselineskip=27pt Quantum field theory which describes the ultimate behavior of elementary
particles and fields in physics fundamentally depends on the concepts of the
bosonic oscillator described by the algebraic relation 
\begin{eqnarray}
aa^{\ast }-a^{\ast }a=1
\end{eqnarray}

\noindent and the fermionic oscillator described by the algebraic relations
\begin{eqnarray}
cc^{\ast }+c^{\ast }c=1 \\ 
c^{2}=0.
\end{eqnarray}

\noindent The algebra of the bosonic oscillator $(1)$ is invariant under the
inhomogeneous symplectic group $ISp(2,R)$ which transforms $a$, $a^{\ast }$
and 1 into each other.The homogenous part of this group which just
transforms $a$, $a^{\ast }$ into each other is $Sp(2,R)$ $\simeq SU(1,1)$%
. For the fermionic oscillator $(2)$\ we should like to remark the
importance of the relation $c^{2}=0.$ Fermions satisfy the Pauli Exclusion
Principle, the two identical fermions can not occupy the same state. Thus $%
c^{2}=0$. The algebra $(2)$ describes the simplest nontrivial quantum
mechanical system. In this sense it is the most fundamental. Although the
bosonic oscillator $(1)$ has a classical limit in which it reduces to the
harmonic oscillator, the fermionic oscillator $(2)$ has no classical
analogue. Thus a thorough understanding of all its properties is important.
One important property of the algebra $(2)$ is that it does not admit a
q-deformation $[1-4]$. Another is that although it is invariant under the
orthogonal group $O(2)$ which transforms $c$ and $c^{\ast }$ into each other
there is no inhomogeneous classical Lie group which transforms $c,c^{\ast }$
and $1$ into each other. In this paper we construct a quantum group $[5-8]$
which achieves this purpose. We show that the structure of the inhomogeneous
invariance group of the fermionic oscillator is a quantum group, that is,
the matrix elements of the transformation matrix which transforms $c,$ $%
c^{\ast }$ and $1$ into each other belong to a noncommutative Hopf algebra $%
[5-8]$ where the coproduct is given by the matrix product. We will develop
the R-matrix formulation of this quantum group and show that the
operators generating this quantum group have a two dimensional
representation which we explicitly construct. The representation matrices
depend on five parameters. We finally present a discussion of our results.

To show that the structure of the inhomogeneous invariance `group' of the
fermionic oscillator is a quantum group, we consider a $3\times 3$ matrix A
whose elements belong to an algebra { $A$}. We form the column matrix
\begin{eqnarray}
{\bf c}=\left[\begin{array}{c}
c \\
c^{\ast } \\ 
1
\end{array}
\right]\end{eqnarray}

\noindent and assume that the action of the matrix A on $c$ is given by
\begin{eqnarray}
\ \ {\bf c}^{^{\prime }}=A\stackrel{\bullet }{\otimes }{\bf c}
\end{eqnarray}

\noindent  We assume that matrix A is of the form
\begin{eqnarray}
A=\left[ \begin{array}{ccc}
\alpha & \beta & \gamma \\ 
\beta ^{\ast } & \alpha ^{\ast } & \gamma ^{\ast } \\ 
0 & 0 & 1
\end{array}
\right] \end{eqnarray}

\noindent in accordance with the general form of inhomogeneous transformations of $c$
and $c^{\ast }$ so that the transformed fermion algebra generators in (4)
are explicitly given by
\begin{eqnarray}
c^{^{\prime }}=\alpha \otimes c+\beta \otimes c^{\ast }+\gamma \otimes 1 \\ 
c^{\ast ^{\prime }}=\alpha ^{\ast }\otimes c^{\ast }+\beta ^{\ast }\otimes
c+\gamma ^{\ast }\otimes 1.
\end{eqnarray}

\noindent If $\alpha ,$ $\beta ,\gamma $ are taken as complex
numbers the invariance of the fermion algebra $(2)$ requires that $\alpha
=\beta =0,$ $\alpha =e^{i\rho }$ or $\gamma $ $=\alpha =0$ $,\beta
=e^{-i\rho }$ giving the homogeneous group $O(2)$. If we assume that $\alpha ,$
$\beta ,$ $\gamma $\ and their hermitian conjugates can form a
noncommutative algebra, the conditions that ${\bf c}^{^{\prime }}$ satisfies
the relations $(2)$ give rise to $12$ (real) relations. It is known that
when some nonlinear completely integrable systems are quantized the Lie
group which describes the symmetries of the system has also to be quantized
to yield a Hopf algebra. We thus look for a Hopf algebra structure for 
{$A$}. In order to accomplish this, these $12$ relations must be
supplemented by $5$ additional relations so that the $17$ defining relations
of the algebra {$A$} are given by
\begin{eqnarray}\alpha \alpha ^{\ast }=\alpha ^{\ast }\alpha \\ 
\beta \beta ^{\ast }=\beta ^{\ast }\beta \\ 
\gamma \gamma ^{\ast }+\gamma ^{\ast }\gamma =1-\alpha ^{\ast }\alpha -\beta
^{\ast }\beta \\ 
\alpha \beta =\beta \alpha \\ 
\alpha \beta ^{\ast }=\beta ^{\ast }\alpha \\ 
\gamma ^{2}=-\alpha \beta \\ 
\alpha \gamma =-\gamma \alpha \\ 
\alpha \gamma ^{\ast }=-\gamma ^{\ast }\alpha \\ 
\beta \gamma =-\gamma \beta \\ 
\beta \gamma ^{\ast }=-\gamma ^{\ast }\beta \ 
\end{eqnarray}

\noindent plus the hermitian conjugates of $(12-18)$. Note
that $\alpha ,$ $\alpha ^{\ast },$ $\beta ,$ $\beta ^{\ast },$ $1$ commute
among themselves. For the special case $\alpha $ $=\beta =0,$ $\gamma ,$ $%
\gamma ^{\ast }$ and $1$ satisfy the fermion algebra. We find the Hopf
algebra with the coproduct
given by matrix multiplication
\begin{eqnarray}\Delta (A)=\left[ 
\begin{array}{ccc}
\Delta (\alpha ) & \Delta (\beta ) & \Delta (\gamma ) \\ 
\Delta (\beta ^{\ast }) & \Delta (\alpha ^{\ast }) & \Delta (\gamma ^{\ast })
\\ 
0 & 0 & 1
\end{array}
\right] =A  \stackrel{\bullet }{\otimes }A,
\end{eqnarray}

\noindent the counit given by the unit matrix
\begin{eqnarray}
\ \epsilon (A)=I
\end{eqnarray}

\noindent  and the antipode given by
\begin{eqnarray}
S(A)= \delta ^{-1}\left[ 
\begin{array}{ccc}
\alpha ^{\ast } & -\beta & -\alpha ^{\ast }\gamma +\beta \gamma ^{\ast } \\ 
-\beta ^{\ast } & \alpha & -\alpha \gamma ^{\ast }+\beta ^{\ast }\gamma \\ 
0 & 0 & 1
\end{array}
\right] 
\end{eqnarray}

\noindent where
\begin{eqnarray} \delta =\alpha \alpha ^{\ast }-\beta \beta ^{\ast }
\end{eqnarray}

\noindent is a central element of the algebra.

The fermionic oscillator algebra $(2-3)$ can be written as a vector algebra 

 \begin{eqnarray}
R{\bf C}_{1}{\bf C}_{2}{\bf =C}_{2}{\bf C}_{1}
\end{eqnarray}

\noindent where

\begin{eqnarray} {\bf C}%
_{1}{\bf C}_{2}=\left[ 
\begin{array}{c}
c^{2} \\ 
cc^{\ast } \\ 
c \\ 
c^{\ast }c \\ 
(c^{\ast })^{2} \\ 
c^{\ast } \\ 
c \\ 
c^{\ast } \\ 
1
\end{array}
\right] ,\hspace{10pt}  {\bf C}_{2}{\bf C}_{1}=\left[ 
\begin{array}{c}
c^{2} \\ 
c^{\ast }c \\ 
c \\ 
cc^{\ast } \\ 
(c^{\ast })^{2} \\ 
c^{\ast } \\ 
c \\ 
c^{\ast } \\ 
1
\end{array}
\right] 
\end{eqnarray}

\noindent and the $9\times 9$ R-matrix is 

\begin{eqnarray} R=\left[ 
\begin{array}{ccccccccc}
-1 & 0 & 0 & 0 & 0 & 0 & 0 & 0 & 0 \\ 
0 & -1 & 0 & 0 & 0 & 0 & 0 & 0 & 1 \\ 
0 & 0 & 1 & 0 & 0 & 0 & 0 & 0 & 0 \\ 
0 & 0 & 0 & -1 & 0 & 0 & 0 & 0 & 1 \\ 
0 & 0 & 0 & 0 & -1 & 0 & 0 & 0 & 0 \\ 
0 & 0 & 0 & 0 & 0 & 1 & 0 & 0 & 0 \\ 
0 & 0 & 0 & 0 & 0 & 0 & 1 & 0 & 0 \\ 
0 & 0 & 0 & 0 & 0 & 0 & 0 & 1 & 0 \\ 
0 & 0 & 0 & 0 & 0 & 0 & 0 & 0 & 1
\end{array}
\right] .
\end{eqnarray}

\noindent The invariance of $(23)$ under the transformation $(5)$ implies
that the matrix A satisfies
\begin{eqnarray}
RA_{1}A_{2}=A_{2}A_{1}R.
\end{eqnarray}

\noindent Since the R-matrix satisfies the quantum Yang-Baxter equation then the
matrix A whose entries satisfy $(9-18)$ defines a quantum matrix group.

The only irreducible representation of the fermion algebra $(2)$ is two
dimensional and can be written in terms of the Pauli matrices
\begin{eqnarray} 
\begin{array}{cc}
c=\frac{1}{2}(\sigma _{1}+i\sigma _{2})=\sigma _{+} \\ 
c^{\ast }=\frac{1}{2}(\sigma _{1}-i\sigma _{2})=\sigma _{-}.
\end{array}
\end{eqnarray}

\noindent  The overall phase $\rho $ can be identified
with the familiar $SO(2)$ group acting on $c$ by $c\rightarrow e^{i\rho }c$.
Since for the special case $\alpha =\beta =0$ representations of the algebra 
{$A$} given by $(9-18)$ must reduce to the representations of the
fermion algebra $(2)$ we may deduce that, if representations of {$A$}
depend on a number of parameters which take special values for the case $%
\alpha =\beta =0$ then {$A$} can only have a two dimensional irreducible
representation. This representation is given by
\begin{eqnarray} 
\begin{array}{ccc}
\alpha =\alpha _{3}\sigma _{3} \\ 
\beta =\beta _{3}\sigma _{3} \\ 
\gamma =\gamma _{+}\sigma_{+}+\gamma _{-}\sigma_{-}
\end{array}
\end{eqnarray}

\noindent  where the complex
numbers $\alpha _{3},$ $\beta _{3},$ $\gamma _{+},$ $\gamma _{_{-}}$ are
chosen such that $(9-18)$ are satisfied
\begin{eqnarray} \begin{array}{c}
\left| \alpha _{3}\right| ^{2}+\left| \beta _{3}\right| ^{2}+\left| \gamma
_{+}\right| ^{2}+\left| \gamma _{-}\right| ^{2}=1 \\ 
\gamma _{+}\gamma _{-}+\alpha _{3}\beta _{3}=0.
\end{array}
\end{eqnarray}

\noindent A particular parametrization is given by
\begin{eqnarray} \begin{array}{c}
\alpha _{3}=e^{i(\rho +\sigma )}\cos \theta \cos \varphi \\ 
\beta _{3}=e^{i(\rho -\sigma )}\sin \theta \sin \varphi \\ 
\gamma _{+}=e^{i(\rho +\tau )}\cos \theta \sin \varphi \\ 
\gamma _{-}=-e^{i(\rho -\tau )}\sin \theta \cos \varphi .
\end{array}
\end{eqnarray}

\noindent Thus the fermionic inhomogeneous orthogonal quantum group $FIO(2)$ depends on $5$ angles. The
fact that the parameters are all angles as well as $(11)$ shows that $FIO(2)$
is compact as compared to the bosonic invariance group $ISp(2)$ which has
the same number of parameters but is noncompact. In contrast, the familiar
inhomogeneous orthogonal group $IO(2)$ which is the Euclidean group in $2$
dimensions has $3$ parameters.

In physics, symmetries are important. Fundamental examples are the Lorentz
invariance of special relativity and the rotational invariance of hydrogen
atom. Until the 1980's it was thought that when a physical system is
quantized the classical group which describes the symmetries of the system
remain intact. i.e it remains a classical group. When some nonlinear,
completely integrable systems were quantized in the 80's it was discovered
that the group which describes the symmetries of the physical system has
also to be quantized, namely the classical group which acts on the classical
system has to change into a quantum group. In the simplest examples, the
classical matrix with commuting elements has to turn into a matrix with
noncommuting elements. Moreover the algebra generated by these elements has
to satisfy the axioms of a Hopf algebra. The quantum groups discovered in
this fashion have the property that in some limit they reduce to a classical
group. A fermionic system, on the other hand, does not have any classical
analogue. By showing that the 'inhomogeneous invariance group' of the
fermionic oscillator is not a classical group but a quantum group, we have
remotivated the introduction of quantum groups into field theory.

\pagebreak
\bigskip

{\bf REFERENCES}

\begin{description}
\item {$^1$} Ws Chung, Internat\i onal Journal
of Theoretical Journal of Theoretical Physics 33, 1611 (1994).

\item {$^2$} A. Solomon, R. McDermot, J. Phys.A: Math. Gen 27, 2619 (1994).

\item {$^3$} Ws Chung, Prog. Theor. Physics 95, 3, (1996).

\item {$^4$} An-Min Wang, Si-Cong Jing and Tu-Nan Ruan, Il Nuovo C\i mento 107A,
9 (1994).

\item {$^5$} L.D. Faddeev, N.Y. Reshetikhin, and L.A. Takhtajan, Algebraic AnalysisM
1, 129 (1988).

\item {$^6$} M. Jimbo, Lett. Math. Phys. 11, 247 (1986).

\item {$^7$} V.G.Drinfeld, Proc. Int. Congr. Math., Berkeley 1, 798 (1986).

\item {$^8$} S.L. Woronowicz, Commun. Math. Phys. 111, 613 (1987).

\end{description}

\end{document}